\documentclass[aps,reprint,twocolumns,prl,superscriptaddress,floatfix,notitlepage,nofootinbib]{revtex4-1}
\usepackage{amssymb,amsmath,amsfonts} 
\usepackage{mathtools}
\usepackage{physics}
\usepackage{array}
\usepackage{graphicx,epsfig,xcolor}
\usepackage[colorlinks=true,citecolor=blue,linkcolor=red,pdftitle={TheoryDynamicalPhaseTransitionsSymmetryBreaking},pdfauthor={Corps and Relano}]{hyperref}
\usepackage{newtxtext,newtxmath}

\newcommand{\hatmath}[1]{\hat{\mathcal{#1}}} 

\begin{document}

\title{Theory of dynamical phase transitions in quantum systems with symmetry-breaking eigenstates}

\author{\'{A}ngel L. Corps}
    \email[]{corps.angel.l@gmail.com}
    \affiliation{Instituto de Estructura de la Materia, IEM-CSIC, Serrano 123, E-28006 Madrid, Spain}
    \affiliation{Grupo Interdisciplinar de Sistemas Complejos (GISC),
Universidad Complutense de Madrid, Av. Complutense s/n, E-28040 Madrid, Spain}
    
\author{Armando Rela\~{n}o}
    \email[]{armando.relano@fis.ucm.es}
    \affiliation{Grupo Interdisciplinar de Sistemas Complejos (GISC),
Universidad Complutense de Madrid, Av. Complutense s/n, E-28040 Madrid, Spain}
    \affiliation{Departamento de Estructura de la Materia, F\'{i}sica T\'{e}rmica y Electr\'{o}nica, Universidad Complutense de Madrid, Av. Complutense s/n, E-28040 Madrid, Spain}

\date{\today} 

\begin{abstract}
We present a theory for the two kinds of dynamical quantum phase transitions, termed DPT-I and DPT-II, based on a minimal set of symmetry assumptions. In the special case of collective systems with infinite-range interactions, both are triggered by excited-state quantum phase transitions. For quenches below the critical energy, the existence of an additional conserved charge, identifying the corresponding phase, allows for a nonzero value of the dynamical order parameter characterizing DPTs-I, and precludes the main mechanism giving rise to nonanalyticities in the return probability, trademark of DPTs-II. We propose a statistical ensemble describing the long-time averages of order parameters in DPTs-I, and provide a theoretical proof for the incompatibility of the main mechanism for DPTs-II with the presence of this additional conserved charge. Our results are numerically illustrated in the fully-connected transverse-field Ising model, which exhibits both kinds of dynamical phase transitions. Finally, we discuss the applicability of our theory to systems with finite-range interactions, where the phenomenology of excited-state quantum phase transitions is absent. We illustrate our findings by means of numerical calculations with experimentally relevant initial states.
\end{abstract}

\maketitle

\textit{Introduction.-} Recent advances in experimental techniques with cold atoms and trapped
ions \cite{Gring2012, Hofferberth2007, Hild2014, Yuznashyan2006,Baumann2010} have stimulated the research about some fundamental
concepts, like dynamical quantum phase transitions (DPTs). DPTs
refer to two distinct phenomena. DPTs-I occur when the dynamics of observables qualitatively changes at a critical value of a control parameter \cite{Eckstein2008,Moeckel2008,Eckstein2009,Sciolla2011,Zhang2017,Muniz2020,Halimeh2017prethermalization}; they are characterized by dynamical order parameters becoming zero at the critical value of the control parameters  \cite{Marino2022}. DPTs-II happen when the dynamics becomes nonanalytic at particular critical times \cite{Heyl2013,Heyl2014,Jurcevic2017,Homrighausen2017,Halimeh2017,Heyl2019,Nicola2021,Halimeh2020,Naji2022,Jafari2019,Jafari2021}; they are purely nonequilibrium phenomena and cannot be described by dynamical order parameters. Both may appear in the same models, like the long-range or fully-connected transverse-field Ising model \cite{Zunkovic2018,Lang2018}, or the Rabi model \cite{Puebla2020}. Although connections between them have been found \cite{Zunkovic2018,Weidinger2017,Lang2018,Hashizume2022,Sehrawat2021,Zunkovic2015,Lerose2019,Lang2018concurrence,Zhou2019}, a common mechanism for their appearance is missing \cite{Heyl2018}.

In this Letter, we present a theory for these two kinds of DPTs. It is inspired in collective systems, where excited-state quantum phase transitions (ESQPTs) \cite{Cejnar2021} entail important consequences for the thermalization after a nonequilibrium process \cite{Puebla2013,Puebla2013b,Puebla2015,Santos2016,Santos2015,Kloc2021}. The corresponding critical energy splits the spectrum into two different phases: one where there exists a constant of motion, $\hat{{\mathcal C}}$, acting like a partial dynamical symmetry, and another where $\hat{{\mathcal C}}$ is no longer constant \cite{Corps2021}. Here, we show that the very existence of this partial dynamical symmetry provides a framework for DPTs, because: (i) the dynamical order parameter characterizing DPTs-I can be different from zero if and only if $\hat{{\mathcal C}}$ is a constant of motion, and (ii) the constancy of $\hat{{\mathcal C}}$ precludes the main mechanism accounting for DPTs-II in systems with a discrete $\mathbb{Z}_2$ symmetry \cite{Heyl2014}. These facts are independent of whether the system exhibits ESQPTs, so our theory applies to both fully-connected and finite-range interacting models. We use the fully-connected transverse-field Ising model \cite{Zunkovic2018,Lang2018} for numerical demonstration, and conclude by discussing the applicability to finite-range interactions systems. 

{\em Assumptions.-} Consider a generic Hamiltonian, $\hat{H}(\lambda)$, depending on a
control parameter, $\lambda$, such that:

(i) There exists a ${\mathbb Z}_2$
symmetry, represented by an operator, $\hat{\Pi}$, which we call parity, $\left[\hat{H}(\lambda),\hat{\Pi} \right]=0$, $\forall
\lambda$, labeling the Hamiltonian eigenstates as $\hat{\Pi}\ket{E_{n,\pm}}=\pm\ket{E_{n,\pm}}$. 

(ii) There exists a second $\mathbb{Z}_{2}$ operator, $\hat{\mathcal{C}}$, acting like a partial symmetry, commuting with the projectors, $\hat{P}_{n}$, onto the energy subspaces only below a given critical energy, $E_{c}$, $[\hat{\mathcal{C}},\hat{P}_{n}]=0$, $\forall E_{n}<E_{c}$. 

(iii) The two previous operators are not commuting, $[\hat{\Pi},\hat{\mathcal{C}}]\neq 0$, so eigenlevels are degenerate below $E_{c}$, and non-degenerate above $E_c$ \cite{Corps2021}.

In the thermodynamic limit (TL), these conditions apply to fully-connected systems exhibiting an ESQPT at $E=E_{c}$ \cite{Corps2021,Corps2022,Corps2022arxiv}, such as the Lipkin-Meshkov-Glick (LMG) \cite{Lipkin1965,Dusuel2004,Heiss2005,Leyvraz2005,Castanos2006,Ribeiro2007,Ribeiro2008,Relano2008,GarciaRamos2017}, the Dicke and Rabi \cite{Puebla2016,Hwang2015,Lobez2016,Bastarrachea2014,Perez2011b,Perez2011,Brandes2013,Kloc2018}, and the two-site Bose-Hubbard models \cite{Relano2014}, spinor Bose-Einstein condensates \cite{Feldmann2021}, and the coupled top \cite{Wang2021}.  As discussed below, they also apply to more general, finite-range interactions systems, where ESQPTs are absent.

{\em Theory for DPTs-I.-} The equilibrium state of a closed quantum system is equal to the infinite-time average of the time-evolved wavefunction \cite{Reimann2008}. Thus, long-time averages inducing dynamical order parameters for DPTs-I can also be described via equilibrium ensembles. Here, we focus on the TL, and therefore assume that all three previous assumptions hold, to build an equilibrium ensemble depending on all relevant constants of motion for the previous setup. Above $E_c$, such ensemble must depend only on energy and parity; it must be diagonal in the parity eigenbasis. Below $E_c$, it must depend also on $\hat{{\mathcal C}}$; as $\left[\hat{\Pi}, \hat{{\mathcal C}} \right] \neq 0$, it must be nondiagonal in the same basis, and therefore it must store information about quantum coherence between parity sectors. As such coherence is complex-valued, a real operator like $\hat{{\mathcal C}}$ \cite{Corps2021} cannot account for all its possible values. Therefore, a third operator is needed:
 $\hat{{\mathcal K}} \equiv  \frac{i}{2} \left[ \hat{{\mathcal C}}, \hat{\Pi} \right].$ It can be shown that $\hat{{\mathcal K}}$ commutes with
the energy projectors below $E_{c}$ if $\hat{{\mathcal
    C}}$ does \cite{Companion}. Hence, such generic system is characterized by a set of three noncommuting charges, $\lbrace \hat{\Pi}, \hat{{\mathcal C}}, \hat{{\mathcal K}} \rbrace$, for $E<E_c$, whereas only $\hat{\Pi}$ is a constant of motion above $E_c$.
This seemingly implies that we need two equilibrium ensembles to describe our setup, and that the transient region around $E=E_{c}$ is ill-defined. However, we can fix this problem defining two new operators,  $\widetilde{C} = \mathbb{I}_{E<E_c} \hat{{\mathcal C}} \, \mathbb{I}_{E<E_c},$  $\widetilde{K} = \mathbb{I}_{E<E_c} \hat{{\mathcal K}} \, \mathbb{I}_{E<E_c},$  where $\mathbb{I}_{E<E_c} \equiv \sum_{n} \theta_{n} \hat{P}_n$, $\hat{P}_n$
is the projector onto the eigenspace with energy $E_n$, and
$\theta_n=1$ if $E_n<E_c$ and $\theta_n=0$ if
$E_n>E_c$. $\langle \widetilde{C}\rangle $ and $\langle \widetilde{K}\rangle $ are equal to $\langle \hat{{\mathcal C}}\rangle $ and $\langle \hat{{\mathcal K}}\rangle $ below $E_{c}$, but identically zero above $E_{c}$. Thus, $\widetilde{C}$ and $\widetilde{K}$ commute with the Hamiltonian in the TL, so we can build an unique equilibrium ensemble from them. 

The simplest equilibrium ensemble depending on these charges is
\begin{equation}\label{eq:rhogme}
  \rho_{\textrm{GME}} (E, p, c, k) = \rho_\textrm{ME} (E) \left( \mathbb{I} + p \, \hat{\Pi} + c \, \widetilde{C} + k \, \widetilde{K} \right),
\end{equation}
where $\rho_\textrm{ME}(E)$ denotes the standard microcanonical ensemble \cite{Alessio2016}.
Parity doublets, $\ket{E_{n,\pm}}$, within a small
energy window around the average energy, $
\textrm{Tr} [\rho_{\textrm{GME}} \hat{H}]$, are equally populated, and
$\textrm{Tr} \rho_\textrm{GME}(E)=1$. We call $\rho_{\textrm{GME}}(E,p,c,k)$ {\em
generalized microcanonical ensemble} (GME). Besides the average energy, it depends on three parameters $p,c,k\in\mathbb{R},$ satisfying $p^{2}+c^{2}+k^{2}\leq 1$, which are fixed by requiring that $\textrm{Tr} [ \rho_{\textrm{GME}} \hat{\Pi}] = \langle \hat{\Pi} \rangle$, $\textrm{Tr}[ \rho_{\textrm{GME}} \widetilde{C}] = \langle \widetilde{C}\rangle$, and $\textrm{Tr}[ \rho_{\textrm{GME}} \widetilde{K}] = \langle \widetilde{K} \rangle$. Equation \eqref{eq:rhogme} represents an ensemble depending on three noncommuting charges \cite{Guryanova2016,Halpern2016,Halpern2020,Fagotti2014}. It has the following properties:

(a) It accounts for the quantum coherence between parity sectors if and only if $E<E_c$. $\rho_{\textrm{GME}}(E,p,c,k)$ has nondiagonal elements in the parity eigenbasis if $c\neq 0$ and/or $k \neq 0$. Consequently, the expectation value of any parity-changing observable, $\hat{\mathcal{O}}$, $\textrm{Tr} \left[ \rho_{\textrm{GME}}(E,p,c,k)\hat{\mathcal{O}} \right]$, may be nonzero only in this spectral region.

(b) It becomes diagonal when all populated eigenstates are above $E_c$. Hence, if $E>E_c$, $\textrm{Tr}\left[ \rho_{\textrm{GME}}(E,p,c,k) \hat{\mathcal{O}} \right]=0$ for any initial condition.

These two properties imply that {\em a DPT-I happens when a quench crosses the critical energy $E_{c}$}. If the initial state fulfills $E<E_c$ and a quench of control parameters $\lambda_{i}\to\lambda_{f}$ leads the system to a region where all populated states are above $E_c$, then all information about quantum coherence between parity sectors is lost. Eq. \eqref{eq:rhogme} is exact in the TL, when assumption (ii) is exactly fulfilled \cite{note1}. For finite-size effects, see below. 

\textit{Theory for DPTs-II.-} Let us consider a quantum quench $\lambda_{i}\to\lambda_{f}$ where the initial state $\ket{\Psi_{0}(\lambda_{i})}$ is a broken-symmetry ground-state, 
\begin{equation}\label{eq:initialstate}
    \ket{\Psi_{0}(\lambda_{i})}=\sqrt{\omega}\ket{E_{0,+}(\lambda_{i})}+e^{i\phi}\sqrt{1-\omega}\ket{E_{0,-}(\lambda_{i})},
\end{equation}
with $\omega\in[0,1]$ and $\phi\in[0,2\pi)$ \cite{note3}. After the quench, its time evolution is given by $\ket{\Psi_{t}(\lambda_{f})}=e^{-i\hat{H}(\lambda_{f})t}\ket{\Psi_{0}(\lambda_{i})}$. In systems with a discrete $\mathbb{Z}_2$ symmetry, DPTs-II are signaled by nonanalytic points in $\mathcal{L}(t)\equiv \mathcal{L}_+ (t) + \mathcal{L}_-(t)$, where $\mathcal{L}_{\pm} = \left| \langle E_{0,\pm}(\lambda_{i})| \Psi_{t}(\lambda_{f})\rangle \right|^2$
are the return probabilities to the positive-parity and negative-parity projections of the initial state \cite{Heyl2014} after the quench. These nonanalytic points are usually identified through the rate $r_N(t)=-\ln \mathcal{L}(t)/N$, akin to a dynamical version of the intensive equilibrium Helmholtz free energy \cite{Heyl2013}. Generically, it is expected that $\mathcal{L}_{\pm} = \exp[-N \Omega_{\pm} (t)]$ \cite{Heyl2013}, where $\Omega_{\pm}(t)$ is an intensive function \cite{Companion}. Hence, in the TL the term with the smallest $\Omega_{\pm}(t)$ dominates, nonanalyticities appearing at any crossing time when $\Omega_+ (t) = \Omega_-(t)$ \cite{Heyl2014}. As in standard phase transitions, $r_N(t)$ remains analytic for $N<\infty$, only becoming singular when $N\to\infty$. From \eqref{eq:initialstate}, we have
\begin{equation}\label{eq:Lpm}
    \mathcal{L}_{+}(t)=\omega |f_{+}(t)|^{2},\,\,\,\mathcal{L}_{-}(t)=(1-\omega)|f_{-}(t)|^{2},
\end{equation}
with $f_{\pm}(t)=\sum_{n}|c_{n,\pm}|^{2}e^{-iE_{n,\pm}(\lambda_{f})t}$, where the expansion of initial eigenstates $\ket{E_{0,\pm}(\lambda_{i})}=\sum_{n}c_{n,\pm}\ket{E_{n,\pm}(\lambda_{f})}$ allowed by parity conservation has been used.

Suppose that the quench only populates final eigenstates below $E_{c}$. We assume that the gap of states of opposite parity closes exponentially in system size, $E_{n,+}(\lambda_{f})=E_{n,-}(\lambda_{f})+\mathfrak{a}_{n}e^{-\mathfrak{b}_{n}N}$, and that $\left[ \hat{\mathcal{C}}, \hat{P}_n \right]$ also goes exponentially to zero \cite{Corps2021}, implying $\hat{\mathcal{C}}\ket{E_{n,+}(\lambda_{f})}=(\ket{E_{n,-}(\lambda_{f})}+\mathfrak{c}_{n}e^{-\mathfrak{d}_{n}N}\ket{\varphi_{n}})/\sqrt{\mathfrak{N}_{n}}$, where $\bra{\varphi_{n}}\ket{E_{n,-}(\lambda_{f})}=0$ and  $\mathfrak{N}_{n}=1+\mathfrak{c}_{n}^{2}e^{-2\mathfrak{d}_{n}N}$. By expanding $1/\sqrt{\mathfrak{N}_{n}}\simeq 1-\frac{1}{2}\mathfrak{c}_{n}^{2}e^{-2\mathfrak{d}_{n}N}$ and keeping the leading term in the exponential decays, we find $c_{n,+}=\bra{E_{0,+}(\lambda_{i})}\hatmath{C}^{\dagger}\hatmath{C}\ket{E_{n,+}(\lambda_{f})}\simeq c_{n,-}+w_{1}e^{-w_{2}N}c_{n,-}+w_{3}e^{-w_{4}N}$ for some constants $w_{i}$, or, equivalently, $|c_{n,+}|^{2}\simeq |c_{n,-}|^{2}+\textsf{a}_{n}e^{-\textsf{b}_{n}N}$ for some $\textsf{a}_{n}\in\mathbb{C}$, $\textsf{b}_{n}>0$ which may be determined explicitly. Therefore,

\begin{equation}\begin{split}
    f_{+}(t)=\sum_{n}\left(|c_{n,-}|^{2}+\textsf{a}_{n}e^{-\textsf{b}_{n}N}\right)e^{-iE_{n,-}(\lambda_{f})t}e^{-i\mathfrak{a}_{n}e^{-\mathfrak{b}_{n}N}t}.
\end{split}\end{equation}
The first summation is an oscillatory correction to $f_{-}(t)$ reflecting the inexact degeneration of eigenlevels for $N<\infty$; for fixed $N$, such correction is only relevant for exponentially large times $t\sim O(e^{\mathfrak{b}_{n}N})$. This timescale coincides with the relaxation time of the prethermal states represented by the GME; it is huge even for modest system sizes \cite{note2}. The second summation is an overall exponentially damped contribution in $N$ originating from the finite-$N$ matrix element of $\hat{\mathcal{C}}$, also negligible for small $N$. So, in the $N\to\infty$ limit, we have

{\bf Result:} If assumptions (i)-(iii) hold, then $f_+(t) = f_-(t)$, $\forall t$, if $E<E_c$.

{\bf Consequence:} the constancy of $\hat{{\mathcal C}}$ if $E<E_c$ implies $\Omega_+(t)$ and $\Omega_-(t)$ cannot cross. Therefore {\em the main mechanism for DPTs-II is forbidden for quenches below the critical energy and can only happen if the quench leads the system to $E>E_c$.} This follows from simple algebraic manipulations of \eqref{eq:Lpm} \cite{Companion}.

\textit{Numerical results.-} To test our theory, we choose the long-range transverse-field Ising model ($\hbar =1$),
\begin{equation}\label{eq:lipkin}
    \hat{H}(\alpha,\lambda)=-\frac{\lambda}{4 \mathcal{N}(\alpha)}\sum_{i\neq j}^{N}\frac{1}{|i-j|^{\alpha}}\hat{\sigma}_{i}^{x}\hat{\sigma}_{j}^{x}+\frac{h}{2}\sum_{i=1}^{N}\hat{\sigma}_{i}^{z}.
\end{equation}
Here, $\mathcal{N}(\alpha)=\frac{2}{N-1}\sum_{m=1}^{N}\frac{N-m}{m^{\alpha}}$ is the Kac factor \cite{Kac1963}, $N$ is the number of spins, and $\hat{\sigma}^{x,y,z}$ are the Pauli matrices.

We start with the fully-connected case, $\alpha=0$. As $[\hat{H}(0,\lambda),\hat{\mathbf{J}}^{2}]=0$, where $\hat{J}_{x,y,z}=\frac{1}{2}\sum_{i=1}^{N}\hat{\sigma}_{i}^{x,y,z}$, we separate the Hamiltonian matrix in symmetry sectors according to its eigenvalues $j(j+1)$; we focus on the maximally symmetric sector, $j=N/2$. We fix $h=1$, and consider $\lambda$ as the control parameter. This is equivalent to fixing $\lambda$ and varying $h$ \cite{Zhang2017,Muniz2020,Jurcevic2017}, but allows us to identify the critical points in simpler terms. In experiments \cite{Muniz2020}, $h\sim$ MHz; thus, $t\sim $ $\mu$s. 

This model has a QPT at $\lambda_{c}=h$ \cite{Botet1982,Companion}, and an ESQPT at $\epsilon_c \equiv E_{c}/j=-h$ if $\lambda>\lambda_c$ \cite{Dusuel2004,Heiss2005,Leyvraz2005,Castanos2006,Ribeiro2007,Ribeiro2008,Relano2008,Companion}. For $\lambda>\lambda_{c}$ and $E<E_{c}$, its eigenstates become degenerate in parity, 
$\hat{\Pi}\equiv \prod_j \hat{\sigma}^z_j$. This entails that $\hat{\mathcal{C}}=\textrm{sign} \left(\frac{1}{2} \sum_{i=1}^N \hat{\sigma}_i^x \right)$, \textit{which represents the sign of the ferromagnetic order parameter in its extensive form}, is a conserved charge. If $\lambda<\lambda_{c}$, or $\lambda>\lambda_{c}$ and $E>E_{c}$, the $\mathbb{Z}_{2}$ symmetry is restored.

\begin{center}
\begin{figure}[h!]
\hspace*{-0.52cm}\includegraphics[width=0.54\textwidth]{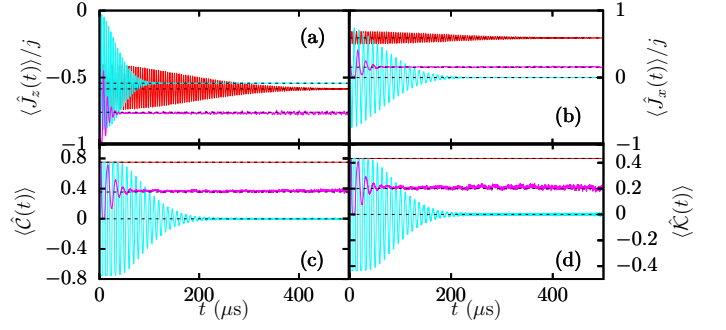}
\caption{Time evolution of physical observables after a quench $\lambda_{i}\to\lambda_{f}=1.75$. The values of the initial control parameter, quenched state average energy and width are $\lambda_{i}=2.5$, $\epsilon(\lambda_{f})=-1.135$, $\sigma_{\epsilon}=0.0025$ (red); $\lambda_{i}=7$, $\epsilon(\lambda_{f})=-1$, $\sigma_{\epsilon}=0.0066$ (magenta), and $\lambda_{i}=27.5$, $\epsilon(\lambda_{f})=-0.89567$, $\sigma_\epsilon=0.0085$ (blue). Black dashed lines show the GME predictions, whose energy window is $[\epsilon(\lambda_{f})-2\sigma_{\epsilon},\epsilon(\lambda_{f})+2\sigma_{\epsilon}]$. System size is $j=6400$; the initial state, \eqref{eq:initialstate}, has $\omega=3/4,\phi=\pi/6$.}
\label{fig:evo}
\end{figure}
\end{center}

In Fig. \ref{fig:evo} we focus on three different quenches $\lambda_{i}\to \lambda_f=1.75$. We compare the time evolution $\langle \hat{\mathcal{O}}(t)\rangle=\bra{\Psi_{t}(\lambda_{f})}\hat{\mathcal{O}}\ket{\Psi_{t}(\lambda_{f})}$ with $\hat{\mathcal{O}}=\hat{J}_{z}, \hat{J}_{x},\hat{\mathcal{C}},\hat{\mathcal{K}}=\frac{i}{2} \left[ \hat{\mathcal{C}},\hat{\Pi} \right]$ with the corresponding GME theoretical predictions. The GME provides a perfect description of long-time averages in all cases. For $\langle \hat{J}_z (t) \rangle$ [Fig. \ref{fig:evo}(a)], we see no traces of DPTs. The usual dynamical order parameter, $\langle \hat{J}_x (t) \rangle$, is in Fig. \ref{fig:evo}(b) \cite{Zhang2017,Muniz2020,Jurcevic2017}. Its long-time average, $\overline{\langle \hat{J}_{x}\rangle}\equiv \lim_{\tau\to\infty}(1/\tau)\int_{0}^{\tau}\textrm{d}t\,\langle\hat{J}_{x}(t)\rangle$, is only zero above $E_c$ (blue). Below this energy, $\langle \hat{J}_x (t)\rangle$ relaxes towards a nonzero value (red), and this also happens when the quench leads the system to the critical energy of the ESQPT (magenta). These observations are easily explained by the GME. As shown in Fig. \ref{fig:evo}(c-d), $\overline{\langle \hat{{\mathcal C}} \rangle}$ and $\overline{\langle \hat{{\mathcal K}}\rangle}$ are nonzero for the first two quenches, meaning that $c\neq0$ and $k\neq0$ in both cases. Note that these two observables are constant for $E<E_c$. At the critical quench, states both below and above $E_{c}$ are populated, so it is possible to find symmetry-breaking long-time averages with $c\neq0$ and $k\neq0$, and also $ \hat{{\mathcal C}} $ and $ \hat{{\mathcal K}}$ are not constant. Finally, $\overline{\langle \hat{{\mathcal C}} \rangle}=\overline{\langle \hat{{\mathcal K}}\rangle}=\overline{\langle \hat{J}_x \rangle}=0$ for the last quench, because $\hat{\mathcal{C}}$ is not constant when $E>E_{c}$. 

\begin{center}
\begin{figure}[h!]
\hspace*{-0.4cm}\includegraphics[width=0.54\textwidth]{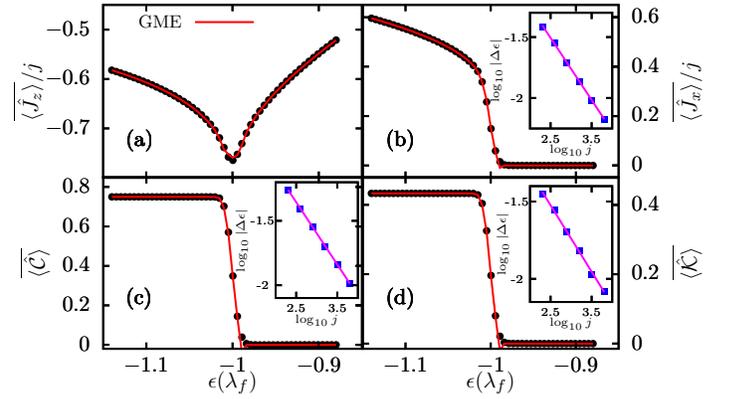}
\caption{(a-d) Long-time average of physical observables after a quench $\lambda_{i}\to\lambda_{f}=1.75$ as a function of the final energy $\epsilon(\lambda_{f})$. $\lambda_{i}$ is varied continuously. System size is $j=6400$; the initial state, \eqref{eq:initialstate}, has $\omega=3/4,\phi=\pi/6$. Points represent exact averages while lines show the GME prediction. Insets in (b-d) show the scaling of $\Delta \epsilon(j)=\epsilon_{c}(j)-\epsilon_{c}(\infty)$ with system size ($\gamma=1/20$), $|\Delta \epsilon(j)|\sim j^{-\kappa},\,\kappa>0.$  }
\label{fig:micros}
\end{figure}
\end{center}

In Fig. \ref{fig:micros} we display the long-time averages and the GME predictions, for quenches with different (average) energies. In Fig. \ref{fig:micros}(a), we observe a minimum of $\overline{\langle \hat{J}_z  \rangle}$ at $E_c$. Yet, the critical behavior is best observed in $\overline{\langle \hat{J}_x \rangle}$, $\overline{\langle \hat{{\mathcal C}}\rangle}$ and $\overline{\langle \hat{{\mathcal K}}\rangle}$: these are nonzero only below $E_{c}$. Similar nonanalytical points should appear also in the inverse participation ratio across the transition \cite{Zhou2019}. The insets of Fig. \ref{fig:micros}(b-d) show how the largest energy, $\epsilon_{c}(j)$, for which $\overline{\langle \hat{J}_x \rangle}$, $\overline{\langle \hat{{\mathcal C}} \rangle}$ and $\overline{\langle \hat{{\mathcal K}}\rangle}$ remain larger than a given bound, $\gamma=1/20$, scales as a function of system size, strongly suggesting that long-time averages vanish exactly at $\epsilon_{c}$ in the TL, as predicted by our theory. Furthermore, the GME provides a perfect description of these equilibrium values.

\begin{center}
\begin{figure}[h!]
\hspace*{-1.2cm}\includegraphics[width=0.6\textwidth]{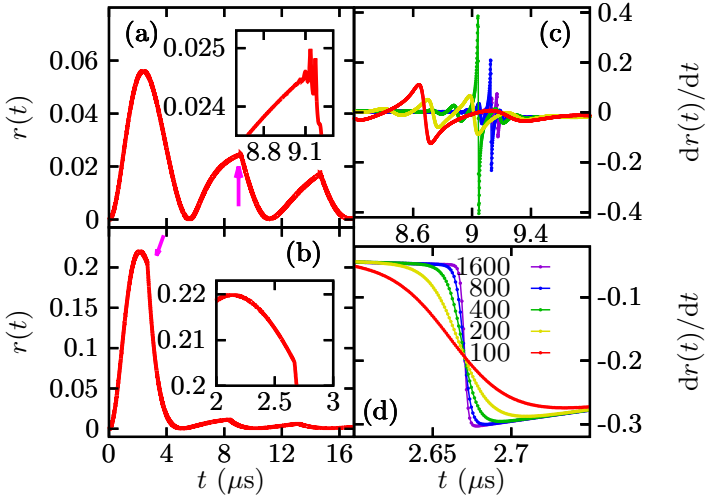}
\caption{(a-b) $r_N(t)$ for quench (a) $\lambda_i=2.535\to\lambda_f=1.6$, ending below $E_{c}$ [$\epsilon(\lambda_{f})=-1.070077$, $\sigma_{\epsilon}=0.00625$] and (b) $\lambda_i=7.437\to\lambda_f=1.6$, ending above $E_{c}$ [$\epsilon(\lambda_{f})=-0.920036$, $\sigma_{\epsilon}=0.0181132$], for $j=1600$; (c-d) finite-size scaling of $\textrm{d}r_{N}(t)/\textrm{d}t$ around the kinks shown by magenta arrows in (a-b), respectively, for several $j$ [legend in (d)]. The initial state \eqref{eq:initialstate} has $\omega=1/2$, $\phi=0$.}
\label{fig:dptii}
\end{figure}
\end{center}

We focus on DPTs-II in Fig. \ref{fig:dptii}. We display $r_N(t)$ for two quenches, one for $E<E_{c}$ [Fig. \ref{fig:dptii}(a)] and another for $E>E_{c}$ [Fig. \ref{fig:dptii}(b)]. At first sight, there seem to exist nonanalytical points in the two cases, those in (a) related to an anomalous dynamical phase \cite{Homrighausen2017,Halimeh2017,Halimeh2020}. To delve into this preliminary conclusion, we display $\textrm{d}r_N(t)/\textrm{d}t$ in Fig. \ref{fig:dptii}(c-d). The finite-size scaling in Fig. \ref{fig:dptii}(c), corresponding to the inset of Fig. \ref{fig:dptii}(a), is inconclusive about the nonanalyticity in the TL: we cannot find a clear pattern as system size increases. Contrarily, Fig. \ref{fig:dptii}(d) shows that $\textrm{d}r_N(t)/\textrm{d}t$ approaches a break as $N$ grows for the kink shown in the inset of Fig. \ref{fig:dptii}(b). Also, all finite-size curves cross at a finite-size precursor of the critical time, $t_c \approx 2.68$. This behavior is reminiscent of first-order phase transitions.

These results can be interpreted with our theory. Below the critical energy, $E<E_c$, $\hat{{\mathcal C}}$ is a conserved charge, in the TL. Therefore, the behavior shown in the inset of Fig. \ref{fig:dptii}(a) cannot be caused by the crossing of $\Omega_+(t)$ and $\Omega_-(t)$. This is the main signature of an anomalous dynamical phase, and its origin and behavior in the TL remain unclear \cite{Companion}. However, above the critical energy, $E>E_c$, $\hat{{\mathcal C}}$ is not a conserved charge, and $\Omega_+(t)$ and $\Omega_-(t)$ may cross. Thus, $r_N(t)$ is expected to become nonanalytic at the corresponding crossing times in the TL, as depicted in Fig. \ref{fig:dptii}(d). Thus, the seemingly nonanalytical precursors appearing in (a,b) have a different origin, manifesting in a completely different scaling to the TL, with no signatures of a phase transition in (c). 

\textit{Beyond fully-connected systems.-} Lastly, we discuss the applicability of our theory to the finite-range transverse-field Ising model, \eqref{eq:lipkin}, with $\alpha=1.2$ (the Kac factor is omitted), and open boundary conditions. We restrict ourselves to the fully symmetric sector regarding the inversion symmetry \cite{Russomanno2021}. This model shows a thermal phase transition for $\alpha \leq 2$ \cite{Dutta2001,GonzalezLazo2021}, and its eigenlevels are degenerate in symmetry pairs below a certain energy, at least if $\alpha \leq 1.5$ \cite{Russomanno2021}.

\begin{center}
\begin{figure}[h!]
\hspace*{-0.7cm}\includegraphics[width=0.54\textwidth]{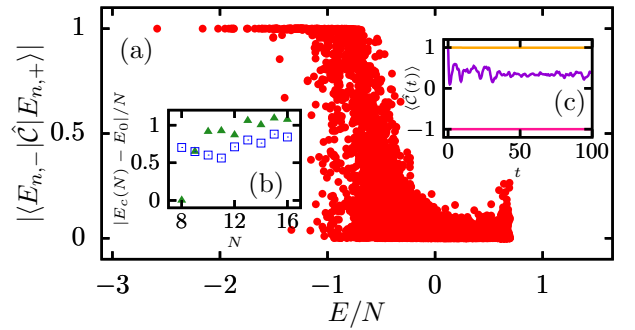}
\caption{(a) Expectation value of $\hat{\mathcal{C}}$ in the eigenstates of opposite parity for \eqref{eq:lipkin}, $\alpha=1.2$, $\lambda=2.5$, $h=1$, $N=16$. (b) Scaling of the precursor of the critical energy $E_{c}(N)$, obtained as the largest eigenvalue such that $1-|\bra{E_{n,-}}\hat{\mathcal{C}}\ket{E_{n,+}}|<10^{-4}$ (triangles) or $|E_{n,+}-E_{n,-}|/\langle s\rangle<10^{-4}$, being $\langle s\rangle$ the mean eigenlevel spacing (squares). (c) Time evolution $\langle \hat{\mathcal{C}}(t)\rangle$ ($N=15$) for clustering initial states and average energies: $\ket{\uparrow\uparrow\uparrow\uparrow\uparrow\uparrow\uparrow\downarrow\uparrow\uparrow\uparrow\uparrow\uparrow\uparrow\uparrow}_{x}$,  $|\langle E\rangle-E_{0}|/N=0.78$ (orange), $\ket{\downarrow \downarrow \downarrow \downarrow \downarrow \downarrow \downarrow \downarrow \downarrow \downarrow \downarrow \downarrow \downarrow \downarrow \downarrow}_{x}$, $|\langle E\rangle-E_{0}|/N=0.026$ (pink), $\ket{\uparrow\uparrow\uparrow\uparrow\downarrow\downarrow\downarrow\downarrow\downarrow\downarrow\downarrow\uparrow\uparrow\uparrow\uparrow}_{x}$, $|\langle E\rangle-E_{0}|/N=1.80$ (purple). }
\label{fig:finiterange}
\end{figure}
\end{center}

In Fig. \ref{fig:finiterange} we focus on $\hat{\mathcal{C}}$, which is defined exactly as before. We observe that $\left| \left\langle E_{n,-} \right| \hat{\mathcal{C}} \left| E_{n,+} \right\rangle \right| \approx 1$ below $E/N \sim -1.5$, which, together with the parity degeneracies \cite{Russomanno2021}, implies the constancy of $\hat{\mathcal{C}}$ in this energy region \cite{Corps2021}. Thus, our assumption (ii) is fulfilled,
the GME is needed to describe steady states within this region, and the main mechanism for DPTs-II is forbidden. Inset (b) shows that two precursors of the {\em critical} energy, obtained from the properties of $\hat{\mathcal{C}}$ (triangles) and the parity degeneracies in the spectrum (squares), behave similarly, becoming stable as the system size increases up to $N=16$. This suggests that our theory should hold below a certain critical energy, $E_c$, in the TL, probably linked to the critical temperature \cite{GonzalezLazo2021}.

Finally, we explore the dynamical consequences of these spectral properties. We select three experimentally relevant initial states satisfying the cluster decomposition property \cite{Sotiriadis2014,Essler2016} (see caption of Fig. \ref{fig:finiterange} for details), with different population probabilities of the Hamiltonian eigenstates. In Fig. \ref{fig:finiterange}(c) we show that $\left\langle \hat{\mathcal{C}}(t) \right\rangle$ remains constant for the first (second) initial condition, $\left\langle \hat{\mathcal{C}}(t) \right\rangle=1$ ($\left\langle \hat{\mathcal{C}}(t)\right\rangle=-1$) which fulfill $E<E_c$; therefore, equilibrium states are described by \eqref{eq:rhogme} with $c=1$ ($c=-1$), $p=k=0$, and the main mechanism for DPTs-II is forbidden. Contrarily, $\left\langle \hat{\mathcal{C}}(t) \right\rangle$ is not constant for the third case, for which $E>E_c$, and thus DPTs-II are possible. This illustrates the applicability of our theory to more general initial states than \eqref{eq:initialstate}.

These results are compatible with the observation of symmetry-breaking steady states \cite{Zunkovic2018,Zhang2017}, together with anomalous DPTs-II \cite{Homrighausen2017,Halimeh2017} in subcritical quenches, whereas regular DPTs-II are observed in supercritical ones \cite{Jurcevic2017}. They also explain the breakdown of the eigenstate thermalization hypothesis into two branches for symmetry-breaking observables \cite{Fratus2017,Fratus2015} ---each should be labeled by a different eigenvalue of $\hat{\mathcal{C}}$. Furthermore, as the conservation of $\hat{\mathcal{K}}$ and $\hat{\Pi}$ implies the conservation of the quantum coherence between the two symmetry-broken magnetization branches, our theory also accounts for the bimodal structure of the full probability distribution of the ferromagnetic order parameter observed in \cite{Ranabhat2022} for small quenches and values of $\alpha$. More work is needed to clarify their link with domain well dynamics \cite{Tan2021,Lerose2019b,Liu2019}.

\begin{acknowledgments}
We gratefully acknowledge discussions with P. P\'{e}rez-Fern\'{a}ndez and J. Dukelsky. A. L. C. is also thankful to J. Novotn\'{y}, P. Str\'{a}nsk\'{y} and P. Cejnar for discussions and their hospitality at Charles University, Prague, when this work was at an advanced stage. This work has been supported by the Spanish grant PGC-2018-094180-B-I00 funded by Ministerio de Ciencia e Innovaci\'{o}n/Agencia Estatal de Investigaci\'{o}n MCIN/AEI/10.13039/501100011033 and FEDER "A Way of Making Europe". A. L. C. acknowledges financial support from `la Caixa' Foundation (ID 100010434) through the fellowship LCF/BQ/DR21/11880024.
\end{acknowledgments}


\begin{thebibliography}{100}
\bibitem{note1} Assumptions (i) and (iii) are also fulfilled for any finite system size.
\end{thebibliography}
\end{document}